\documentstyle[12pt,epsfig]{article}

\catcode`\@=11
\long\def\@makefntext#1{
\protect\noindent \hbox to 3.2pt {\hskip-.9pt  
$^{{\ninerm\@thefnmark}}$\hfil}#1\hfill}                %CAN BE USED 

\def\@makefnmark{\hbox to 0pt{$^{\@thefnmark}$\hss}}  %ORIGINAL 
        
\def\ps@myheadings{\let\@mkboth\@gobbletwo
\def\@oddhead{\hbox{}
\rightmark\hfil\ninerm\thepage}   
\def\@oddfoot{}\def\@evenhead{\ninerm\thepage\hfil
\leftmark\hbox{}}\def\@evenfoot{}
\def\sectionmark##1{}\def\subsectionmark##1{}}

\renewcommand{\thefootnote}{\fnsymbol{footnote}}

\newcounter{sectionc}\newcounter{subsectionc}\newcounter{subsubsectionc}
\renewcommand{\section}[1] {\vspace*{0.6cm}\addtocounter{sectionc}{1} 
\setcounter{subsectionc}{0}\setcounter{subsubsectionc}{0}\noindent 
        {\normalsize\bf\thesectionc. #1}\par\vspace*{0.4cm}}
\renewcommand{\subsection}[1] {\vspace*{0.6cm}\addtocounter{subsectionc}{1} 
        \setcounter{subsubsectionc}{0}\noindent 
        {\normalsize\it\thesectionc.\thesubsectionc. #1}\par\vspace*{0.4cm}}
\renewcommand{\subsubsection}[1]
{\vspace*{0.6cm}\addtocounter{subsubsectionc}{1}
        \noindent {\normalsize\rm\thesectionc.\thesubsectionc.\thesubsubsectionc. 
        #1}\par\vspace*{0.4cm}}

\newcounter{appendixc}
\newcounter{subappendixc}[appendixc]
\newcounter{subsubappendixc}[subappendixc]

\renewcommand{\appendix}[1] {\vspace*{0.6cm}
        \refstepcounter{appendixc}
        \setcounter{figure}{0}
        \setcounter{table}{0}
        \setcounter{equation}{0}
        \renewcommand{\thefigure}{\Alph{appendixc}.\arabic{figure}}
        \renewcommand{\thetable}{\Alph{appendixc}.\arabic{table}}
        \renewcommand{\theappendixc}{\Alph{appendixc}}
        \renewcommand{\theequation}{\Alph{appendixc}.\arabic{equation}}
        \noindent{\bf Appendix \theappendixc #1}\par\vspace*{0.4cm}}

\def\abstracts#1{{
        \centering{\begin{minipage}{12.2truecm}\footnotesize\baselineskip=12pt\noindent
        \centerline{\footnotesize ABSTRACT}\vspace*{0.3cm}
        \parindent=0pt #1
        \end{minipage}}\par}} 

\renewenvironment{thebibliography}[1]
        {\begin{list}{\arabic{enumi}.}
        {\usecounter{enumi}\setlength{\parsep}{0pt}
\setlength{\leftmargin 1.25cm}{\rightmargin 0pt}
         \setlength{\itemsep}{0pt} \settowidth
        {\labelwidth}{#1.}\sloppy}}{\end{list}}

\newcounter{itemlistc}
\newcounter{romanlistc}
\newcounter{alphlistc}
\newcounter{arabiclistc}

\newcommand{\fcaption}[1]{
        \refstepcounter{figure}
        \setbox\@tempboxa = \hbox{\footnotesize Fig.~\thefigure. #1}
        \ifdim \wd\@tempboxa > 6in
           {\begin{center}
        \parbox{6in}{\footnotesize\baselineskip=12pt Fig.~\thefigure. #1}
            \end{center}}
        \else
             {\begin{center}
             {\footnotesize Fig.~\thefigure. #1}
              \end{center}}
        \fi}

\newcommand{\tcaption}[1]{
        \refstepcounter{table}
        \setbox\@tempboxa = \hbox{\footnotesize Table~\thetable. #1}
        \ifdim \wd\@tempboxa > 6in
           {\begin{center}
        \parbox{6in}{\footnotesize\baselineskip=12pt Table~\thetable. #1}
            \end{center}}
        \else
             {\begin{center}
             {\footnotesize Table~\thetable. #1}
              \end{center}}
        \fi}

\def\@citex[#1]#2{\if@filesw\immediate\write\@auxout
        {\string\citation{#2}}\fi
\def\@citea{}\@cite{\@for\@citeb:=#2\do
        {\@citea\def\@citea{,}\@ifundefined
        {b@\@citeb}{{\bf ?}\@warning
        {Citation `\@citeb' on page \thepage \space undefined}}
        {\csname b@\@citeb\endcsname}}}{#1}}

\newif\if@cghi
\def\cite{\@cghitrue\@ifnextchar [{\@tempswatrue
        \@citex}{\@tempswafalse\@citex[]}}
\def\citelow{\@cghifalse\@ifnextchar [{\@tempswatrue
        \@citex}{\@tempswafalse\@citex[]}}
\def\@cite#1#2{{$\null^{#1}$\if@tempswa\typeout
        {IJCGA warning: optional citation argument 
        ignored: `#2'} \fi}}

 1
 1
 1

\font\ninerm=cmr9

\begin{document}
%----------------------------------------------------------------------
\newcommand{\lattice}{
\multiput(0,0)(10,0){18}{\line(0,1){200}}
\multiput(0,0)(0,10){21}{\line(1,0){170}}
}
%----------------------------------------------------------------------
\newcommand{\ra}{\rightarrow}

\begin{flushright}
  BI-TP 97/40\\ MPI-PhT/97-63\\ WUE-ITP-97-040\\[1ex] 
  {\tt hep-ph/9710327}\\ 
\end{flushright}
\vskip 35pt

\centerline{\normalsize\bf THEORETICAL INTERPRETATIONS}
\baselineskip=16pt 
\centerline{\normalsize\bf OF THE HERA HIGH-$Q^2$
  EVENTS\footnote{Talk presented by R.\ R\"uckl at the Ringberg
    Workshop {\it New Trends in HERA Physics}, May 25 -- 30, 1997,
    Tegernsee, Germany.} } 
\baselineskip=22pt

\centerline{\footnotesize R.\ R\"UCKL}
\baselineskip=13pt
\centerline{\footnotesize\it Institut f\"ur Theoretische Physik,
  Universit\"at W\"urzburg}
\baselineskip=12pt
\centerline{\footnotesize\it D-97074 W\"urzburg, Germany}
%\centerline{\footnotesize E-mail: username@abc.def.gh}
\vspace*{0.3cm}
\centerline{\footnotesize and}
\vspace*{0.3cm}
\centerline{\footnotesize H.\ SPIESBERGER}
\baselineskip=13pt
\centerline{\footnotesize\it Fakult\"at f\"ur Physik, Universit\"at
  Bielefeld, D-33501 Bielefeld, Germany}
\vspace*{0.9cm} 
\abstracts{Theoretical interpretations of the high-$Q^2$ events
  recently observed in deep-inelastic positron-proton scattering 
  at HERA are reviewed. We discuss leptoquarks, squarks with 
  $R$-parity violating couplings, and contact interactions. 
  Bounds from and implications on other experiments are taken
  into consideration.}

\normalsize\baselineskip=15pt
\setcounter{footnote}{0}
\renewcommand{\thefootnote}{\alph{footnote}}
%
%%%%%%%%%%%%%%%%%%%%%%%%%%%%%%%%%%%%%%%%%%%%%%%%%%%%%%%%%%%%%%%%%%%%%%%
\section{The Data}

Both HERA experiments, H1 \cite{H1data} and ZEUS \cite{Zdata}, have
reported the observation of an excess of events in deep-inelastic
positron-proton scattering at large values of Bjorken-$x$ and momentum
transfer $Q^2$, relative to the expectation in the standard model.
Including the new data presented recently at the 1997 Lepton-Photon
Symposium \cite{straub}${}^,$\footnote{For a detailed discussion of
  the 1994-96 data see, for example, Ref.\ \cite{sirois}}~, H1 and
ZEUS each observe 18 neutral current (NC) events at $Q^2 > 1.5 \cdot
10^4$ GeV$^2$, while H1 expects $8.0 \pm 1.2$ and ZEUS about 15
events.  At H1, the excess is concentrated in the rather narrow mass
range 187.5 GeV $\leq M = \sqrt{xs} \leq$ 212.5 GeV where 8 events are
observed with $1.53 \pm 0.29$ expected.  However, in the same region,
ZEUS finds roughly the expected number of events. Conversely, in the
region $x>0.55$, $y = Q^2/M^2 >0.25$ where ZEUS finds 5 events with
$1.51 \pm 0.13$ expected, H1 observes no excess. A surplus of events
is also observed in charged current (CC) scattering, although with
smaller statistical significance.  At $Q^2 > 10^4$ GeV$^2$, H1 and
ZEUS together find 28 events and expect $17.7 \pm 4.3$.

The clustering of the H1 events at a fixed value of $M = \sqrt{xs}$
could indicate the production of a resonance with leptoquark quantum
numbers and mass $M \simeq 200$ GeV. On the other hand, ZEUS has 4
events clustered at a somewhat higher mass $M \simeq 225$ GeV.  Given
the experimental mass resolution of 5 and 9 GeV, respectively, it
appears unlikely that both signals come from a single narrow resonance
\cite{straub,BB}.  Rather, the excess may be a continuum effect
reminiscent of contact interactions.  Although the anomalous number of
events is not large enough to clearly exclude statistical fluctuations
as the origin of the effect, and because of the puzzling differences
among the H1 and ZEUS data, it is important to investigate possible
interpretations within and beyond the standard model.

Leaving aside very small uncertainties in electroweak parameters and
radiative corrections \cite{H1data,Zdata}, the main theoretical
uncertainty on the high-$Q^2$ cross sections in the standard model
come from the structure functions.  The latter are obtained by
extrapolation of measurements at lower $Q^2$ using next-to-leading
order evolution equations.  For presently available parametrizations
the HERA collaborations have estimated an uncertainty of about 7\,\%
from this source \cite{H1data,Zdata}.  Attempts \cite{partons} to add
to the conventional parton densities a new valence component at very
large $x$ but low $Q^2$, and to feed down this enhancement to lower
$x$ by evolution to very high $Q^2$, fail to increase the cross
sections by a sufficient amount because of the constraints put by the
fixed-target data.  Explanations based on a strong intrinsic charm
component \cite{charm} produced by some nonperturbative effect do not
seem to be more successful. In fact, up to this day no standard model
mechanism is known which could explain the observed surplus of events.
Moreover, no sign of a deviation from the perturbative evolution of
structure functions in QCD up to $Q^2 \simeq 10^4$ GeV$^2$ has so far
been found in the data.  Whatever mechanism is responsible for the
HERA anomaly, it must have quite a rapid onset.

Therefore, if the excess is not a statistical fluctuation, it is very
likely produced by some new physics beyond the standard model. Then
the question arises whether one is dealing with a (not necessarily
single) resonance or with a continuum effect. In the following, 
we give a brief overview of the main hypotheses put forward. 

%
%%%%%%%%%%%%%%%%%%%%%%%%%%%%%%%%%%%%%%%%%%%%%%%%%%%%%%%%%%%%%%%%%%%%%%
\section{Leptoquarks}

The most exciting speculation is the one of a possible discovery of a
new particle.  Being supposedly produced as a $s$-channel resonance in
$e^+q$ or $e^+ \bar q$ collisions, this new member of the particle zoo
must be a boson and carry simultaneously lepton and quark quantum
numbers.  Such species are generically called leptoquarks.  In
addition to their couplings to the standard model gauge bosons, they
are usually assumed to have Yukawa-type couplings to lepton-quark
pairs:
\begin{equation}
\lambda_i \phi_{LQ} \bar{l}_i q
~~~{\rm or} ~~~
\lambda_i \phi_{LQ} \bar{l}_i q^c~,
\end{equation}
and similarly for vector LQs. Here, $c$ denotes charge conjugation,
and $i=L,R$ specifies the lepton handedness. 

With the above couplings the resonance cross section in $ep$
scattering is given by
\begin{equation}
\sigma = N \frac{\pi}{4s} \lambda^2 q_f(M^2/s,\mu^2)~,
\end{equation}
where $q_f(x,\mu^2)$ is the density of quarks (or antiquarks) with
flavour $f$ in the proton, and $N = 1\,(2)$ for scalars (vectors). The
relevant scale $\mu$ is expected to be of order $M$.  Obviously,
leptoquarks with fermion number $F=0\,(2)$ are dominantly produced
from valence quarks in $e^+q\,(e^-q)$ fusion.
 
Having only couplings to standard model particles, leptoquarks decay
exclusively to lepton-quark pairs. The partial width per channel is
given by
\begin{equation}
\Gamma = \frac{a}{16\pi} \lambda^2 M = 350\,{\rm MeV}\,a
\left(\frac{\lambda}{e}\right)^2 \left(\frac{M}{200\,{\rm GeV}}\right)~, 
\end{equation}
$a$ being 1 for scalars and 2/3 for vectors. Hence, leptoquarks are
very narrow for masses in the range accessible at HERA, and for
couplings weaker than the electromagnetic coupling strength
$e=\sqrt{4\pi\alpha}$.

%----------------------------------------------------------------------
{\small
\begin{table}[htp]
\begin{center}
\begin{tabular}{|c|c|c|c|c|c||c|c|}
\hline
\multicolumn{2}{|c|}{\rule{0mm}{5mm}$LQ$} & 
  $Q$ & 
     Decay & BR &
    Coupling & 
   Limits & HERA \\
\multicolumn{2}{|c|}{}&
  &
     Mode & $e^{\pm}\,j$ & $\lambda_{L,R}$
       & Ref.\ \cite{Leurer} & estimates \\[1mm]
%
%%%%%%%%%%%%%%%%  1
\hline \rule{0mm}{5mm}
$S_0$ & {\raisebox{1.ex}{$\displaystyle
        \begin{array}{c}\rule{0mm}{5mm}\tilde{d}_R\\ ~ \end{array}$}} &
  $-1/3 $ & 
    $\displaystyle \begin{array}{c} e_L u  \\ \nu_L d\\ e_R u \end{array}$ & 
     {\raisebox{-0.8ex}{$\displaystyle 
     \begin{array}{c} \frac{1}{2} \\ \rule{0mm}{6mm} 1 \end{array}$}} &
    $\displaystyle \begin{array}{c} g_L \\ -g_L\\ g_R  \end{array}$ &  
     {\raisebox{-0.8ex}{$\displaystyle 
     \begin{array}{c} g_L<0.06 \\ \rule{0mm}{6mm} g_R<0.1 \end{array}$}} &
     {\raisebox{-0.8ex}{$\displaystyle 
     \begin{array}{c} 0.40 \\ \rule{0mm}{6mm} 0.28 \end{array}$}} 
\\[1mm]
%
%%%%%%%%%%%%%%%%  2
\hline
\multicolumn{2}{|c|}{\rule{0mm}{5mm}$\tilde{S}_0$} & 
  $-4/3 $      &
     $e_R d$ & 1 &
    $g_R$ &
   $g_R<0.1$    & 0.30
\\[1mm]
%
%%%%%%%%%%%%%%%%  3
\hline
\multicolumn{2}{|c|}{\rule{0mm}{5mm}} &  
  $+2/3$ & 
     $\nu_L u$ & 0 &
    $\sqrt{2}g_L$  & 
                & $-$
\\
\cline{3-6}
\multicolumn{2}{|c|}{$S_1$} &
  $-1/3$  & 
     $\displaystyle \begin{array}{c} \nu_L d \\ e_L u \end{array}$ &  
           $\frac{1}{2}$ &
    $\displaystyle \begin{array}{c} -g_L \\ -g_L \end{array}$ & 
   $g_L<0.09$    & 0.40
\\
\cline{3-6}
\multicolumn{2}{|c|}{\rule{0mm}{5mm}} & 
$-4/3$ 
  & 
     $e_L d$ & 1 & 
    $-\sqrt{2}g_L$ & 
                 & 0.21
\\[1mm]
%
%%%%%%%%%%%%%%%  4
\hline
\multicolumn{2}{|c|}{\rule{0mm}{5mm}} &  
  $-1/3$ &
     $\displaystyle \begin{array}{c} \nu_L d \\ e_R u \end{array}$ & 
     $\displaystyle \begin{array}{c} 0 \\ 1 \end{array}$ & 
    $\displaystyle \begin{array}{c} g_L \\ g_R \end{array}$&
   $g_L<0.09$& 
    $\displaystyle \begin{array}{c} - \\ 0.30 \end{array}$
\\[1mm] 
\cline{3-6}
\multicolumn{2}{|c|}{{\raisebox{ 3ex}[-3ex]{ $V_{1/2}$}}}& 
  $-4/3$  & 
     $\displaystyle \begin{array}{c} e_L d  \\  e_R d \end{array}$ &
        $1$ &
    $\displaystyle \begin{array}{c} g_L \\ g_R\end{array}$&
   $g_R<0.05$& 
    $\displaystyle \begin{array}{c} 0.32 \\ 0.32 \end{array}$
\\ [1mm]
%
%%%%%%%%%%%%%%   5
\hline
\multicolumn{2}{|c|}{\rule{0mm}{5mm}} &   
  $ +2/3$ & 
     $\nu_L u$& 0 & 
    $g_L$ &
            & $-$
\\[1mm]
\cline{3-6}
\multicolumn{2}{|c|}{{\raisebox{ 1.5ex}[-1.5ex]{ $\tilde{V}_{1/2}$}}}&
\rule{0mm}{5mm}
  $ -1/3$ & 
     $ e_L u $& 1&
    $g_L$ &
   {\raisebox{ 1.6ex}[-1.6ex]{$g_L<0.09$}} & 0.32
\\[1mm]
\hline
%%%%%%%%%%%%%%%  6
\hline
\multicolumn{2}{|c|}{\rule{0mm}{5mm}} &  
  $-2/3$ &
     $\displaystyle\begin{array}{c}\nu_L \bar{u}\\e_R \bar{d} \end{array}$ &
     $\displaystyle\begin{array}{c} 0 \\ 1 \end{array}$ &
    $\displaystyle\begin{array}{c} g_L \\ -g_R \end{array}$&
   $g_L<0.1$& 
    $\displaystyle\begin{array}{c} - \\ 0.052 \end{array}$
\\[1mm] 
\cline{3-6}
\multicolumn{2}{|c|}{{\raisebox{ 3ex}[-3ex]{ $S_{1/2}$}}}& 
  $-5/3$  & 
     $\displaystyle\begin{array}{c} e_L \bar{u}\\ e_R \bar{u}\end{array}$&
        1 &
    $\displaystyle\begin{array}{c} g_L \\ g_R\end{array}$&
   $g_R<0.09$& 
    $\displaystyle\begin{array}{c} 0.026 \\ 0.026 \end{array}$
\\ [1mm]
%
%%%%%%%%%%%%%%%  7
\hline
\rule{0mm}{6mm} & $\overline{\tilde{d}}_L$ &  
  $ +1/3$ & 
     $\nu_L \bar{d}$& 0 & 
    $g_L$ &
             & $-$
\\[1mm]
\cline{2-6}
\rule{0mm}{5mm}
{\raisebox{3ex}[-3ex]{ $\tilde{S}_{1/2}$}} & 
$\overline{\tilde{u}}_L$&
  $ -2/3$ &
     $ e_L\bar{d} $& 1 &
    $g_L$ &
   {\raisebox{2.5ex}[-2.5ex]{$g_L<0.1$}}& 0.052
\\[1mm]
%
%
%%%%%%%%%%%%%%   8
\hline
\multicolumn{2}{|c|}{\rule{0mm}{6mm}$V_0$} & 
  $-2/3 $ & 
     $\displaystyle \begin{array}{c} e_L \bar{d}  \\
                     \nu_L \bar{u}\\ e_R \bar{d} \end{array}$ & 
     {\raisebox{-0.8ex}{$\displaystyle 
     \begin{array}{c} \frac{1}{2} \\ \rule{0mm}{6mm} 1 \end{array}$}} &
    $\displaystyle \begin{array}{c} g_L \\ g_L\\ g_R  \end{array}$ &  
     {\raisebox{-0.8ex}{$\displaystyle 
     \begin{array}{c} g_L<0.05 \\ \rule{0mm}{6mm} g_R<0.09 \end{array}$}} &
     {\raisebox{-0.8ex}{$\displaystyle 
     \begin{array}{c} 0.080 \\ \rule{0mm}{6mm} 0.056 \end{array}$}}
\\[1mm]
%
%%%%%%%%%%%%%%  9
\hline
\multicolumn{2}{|c|}{\rule{0mm}{5mm}$\tilde{V}_0$} & 
  $-5/3$      & 
     $e_R \bar{u}$ & 1 &
    $g_R$   &
   $g_R<0.09$& 0.027
\\[1mm]
%
%%%%%%%%%%%%%% 10
\hline
\multicolumn{2}{|c|}{\rule{0mm}{5mm}} &  
  $+1/3$         & 
     $\nu_L \bar{d}$ & 0 &
    $\sqrt{2}g_L$  & 
                     & $-$ 
\\
\cline{3-6}
\multicolumn{2}{|c|}{$V_1$} &
  $-2/3$&
     $\displaystyle\begin{array}{c} e_L \bar{d}\\ \nu_L\bar{u}\end{array}$&
         $\frac{1}{2}$ &
    $\displaystyle\begin{array}{c}\rule{0mm}{5mm} -g_L\\g_L \end{array}$ & 
   $g_L<0.04$& 0.080
\\
\cline{3-6}
\multicolumn{2}{|c|}{\rule{0mm}{5mm}} &  
  $-5/3$ & 
     $e_L \bar{u}$ & 1 &
    $\sqrt{2}g_L$ & 
                    & 0.019
\\[1mm]
\hline
\end{tabular}
\caption{\it Scalar ($S$) and vector ($V$) leptoquarks, and their
  electric charges $Q$, decay modes, branching ratios into charged
  lepton + jet channels, and Yukawa couplings. Given are also the most
  stringent low-energy bounds and the couplings deduced from the
  1994-96 HERA data \protect\cite{krsz}.  Inclusion of the 1997 data
  decrease the couplings by about 15\%.  Using the H1 data alone would
  roughly give the couplings shown above.  Also shown are the possible
  assignments of squarks with $R$-parity violating couplings.}
\label{tabprop}
\end{center}
\end{table}}

%----------------------------------------------------------------

Leptoquarks appear in extensions of the standard model involving
unification, technicolor, compositeness, or $R$-parity violating
supersymmetry.  In the generally adopted framework described in Ref.\ 
\cite{BRW}, the Yukawa couplings are taken to be dimensionless and
$SU(3)\times SU(2)\times U(1)$ symmetric. Moreover, they are assumed
to conserve lepton and baryon number in order to avoid rapid proton
decay, to be non-zero only within one family in order to exclude FCNC
processes beyond the CKM mixing, and chiral in order to avoid the very
strong bounds from leptonic pion decays.  The allowed states can be
classified according to spin, weak isospin and fermion number. They
are summarized in Table \ref{tabprop}.

The leptoquark masses and couplings are constrained by high-energy
data.  Direct searches for leptoquarks have been performed at the
Tevatron, at HERA and at LEP. Recently, both experiments, CDF and D0,
have improved their mass limits for scalar leptoquarks considerably.
D0 excludes first generation leptoquarks with masses below 225 GeV
assuming a branching ratio $B_{eq}=1$ for decays into $e^{\pm}$ and a
jet \cite{D0}, whereas CDF quotes a limit of 213 GeV \cite{cdf} (all
mass limits are at 95\,\% CL). For branching ratios less than one, the
limits are weaker, e.g., $M > 176$ GeV for $B_{eq}=0.5$ \cite{D0}. The
bounds on vector states are even stronger: 298 GeV for $B_{eq}=1$ and
270 GeV for $B_{eq}=0.5$ \cite{vector}. The corresponding bounds on
second and third generation scalar leptoquarks are $M > 184$ GeV for
$B_{\mu q} = 1$ and $M > 98$ GeV for $B_{\tau q} = 1$, respectively
\cite{lqst}.  The above constraints follow from pair-production mainly
by $q\bar q$ annihilation, and are therefore practically independent
of the unknown Yukawa coupling $\lambda$.  In contrast, the mass
limits obtained at HERA depend on $\lambda$ and the quantum numbers
specified in Table \ref{tabprop}. For $\lambda = e$ they range from
207 to 272 GeV \cite{hera}.  These limits are lowered by about 50 GeV
if $\lambda=0.1$. Finally, the most stringent but again
$\lambda$-dependent mass bound at LEP2 comes from the search for
single-leptoquark production and excludes masses below 131 GeV
assuming $\lambda > e$ \cite{leps}. The mass limits from leptoquark
pair production \cite{lepp} roughly reach half of the center-of-mass
energy $\sqrt{s}$, and are thus much weaker than the Tevatron
bounds.

Indirect bounds on Yukawa couplings and masses can be derived from
t/u-channel LQ-exchange in $e^+e^- \to q \bar q$ \cite{eeqq}, and from
low-energy data \cite{Leurer}. From the very recent analysis by OPAL
we infer upper limits on $\lambda$ between 0.2 and 0.7 assuming
$M=200$ GeV. However, the most restrictive bounds come from atomic
parity violation and lepton and quark universality, at least for first
generation leptoquarks and chiral couplings. The maximum allowed
couplings for $M=200$ GeV are given in Tab.\ \ref{tabprop}
\cite{krsz}.

In order to explain the observed excess of high-$Q^2$ events at HERA
by the production and decay of a 200 GeV leptoquark, one roughly needs
$\lambda \simeq e$ for $F=2$ states and $\lambda \simeq e/10$ for
$F=0$.  The factor 10 difference in $\lambda$ simply reflects the
factor 100 difference in the sea and valence quark densities in the
region of $x$ and $Q^2$ where the signal resides.  Similarly, the
coupling of $F=0$ leptoquarks to the $d$ quark has to be two times
larger than the coupling to the $u$ quark in order to compensate the
factor four difference in the corresponding quark densities. These
simple rules of thumb describe the main pattern in the couplings found
in detailed analyses \cite{LQ}, and shown in the last column of Table
\ref{tabprop}.

Whereas the coupling strength $\lambda$ required for $F=0$ leptoquarks
is compatible with all existing bounds, the coupling necessary for
$F=2$ leptoquarks is already excluded by the low-energy constraints,
and also at the borderline of getting in conflict with LEP2 data.
Moreover, with such strong couplings, $F=2$ leptoquarks should have
shown up in $e^-p$ scattering at HERA \cite{H1search}, where they can
be produced off the valence quark component, despite of the low
luminosity of the previous $e^-p$ run.  Since vector leptoquarks
cannot be made responsible for an excess of events at $M \simeq 200$
to 225 GeV because of the high Tevatron mass bounds, only the two
scalar doublets $S_{1/2}$ and $\tilde{S}_{1/2}$ remain from the whole
Table \ref{tabprop} as a possible interpretation. However, also these
solutions have difficulties. Firstly, the Tevatron mass limits require
scalar leptoquarks of the first generation with $M \simeq 200$ GeV to
have branching ratios into $e\,+\,jet$ final states less than about
0.7\footnote{This follows from the D0 limits \cite{D0} alone. An even
  smaller branching ratio is required by the combined D0 and CDF
  bounds.}~, whereas in the simple framework considered in Table
\ref{tabprop}, $S_{1/2}$ and $\tilde{S}_{1/2}$ are expected to have
$B_{eq}=1$. Secondly, the scalar doublets cannot give rise to CC
events.

Thus it seems that the leptoquark interpretation of the HERA
high-$Q^2$ events points to more complicated and maybe more realistic
scenarios.  Several possibilities have been suggested allowing for
$B_{eq} < 1$ and, at the same time, providing CC final states:
$SU(2)\times U(1)$ violating, intergenerational couplings
\cite{ks,agm}, leptoquark mixing \cite{babu}, LQ models with
additional vector-like fermions \cite{hr}, and squarks with $R$-parity
violating couplings \cite{squarks}.

\clearpage
%
%%%%%%%%%%%%%%%%%%%%%%%%%%%%%%%%%%%%%%%%%%%%%%%%%%%%%%%%%%%%%%%%%%%%%%
\section{Squarks}

Supersymmetry with $R$-parity violation provides an attractive
theoretical framework in which squarks can have direct couplings to
lepton-quark pairs, and therefore act as leptoquarks.  However,
because of the usual $R$-parity conserving interactions one naturally
expects the branching ratio for $\tilde{q} \to e\,+\,jet$ to be
smaller than unity.  In addition, one can get CC-like final states,
e.g., through the decay chain $\tilde{q} \to q \chi \to q \nu +
\cdots$, $\chi$ being either a neutralino or chargino.  Thus it
appears possible to avoid the two main problems encountered in the
simplest leptoquark models.

In the minimal supersymmetric extension of the standard model, one can
have a renormalizable, gauge invariant operator in the superpotential
that couples squarks to quarks and leptons:
\begin{equation}
W_{\not R} = \lambda'_{ijk} L_L^i Q_L^j \overline{D}_R^k.
\end{equation}
Here, $L$ and $Q$ denote doublets of lepton and quark superfields,
respectively, $D$ stands for singlets of $d$-quark superfields, and
$i$, $j$, and $k$ are generation indices.  This interaction term
violates global invariance of $R$-parity, defined as $R
=(-1)^{3B+L+2S}$ which is $+1$ for particles and $-1$ for
superpartners. In general, there are other $R$-odd operators in the
superpotential that couple sleptons to leptons and squarks to quarks.
Together, they may induce rapid proton decay. This can be avoided by
requiring conservation of $R$-parity, or a strong hierarchy in the
various couplings. Generally, these two options lead to very different
phenomenology.

%----------------------------------
\begin{table}[bh]
\begin{center}
\begin{tabular}{|c|c|c|c|}\hline \rule{0mm}{5mm}
$ijk$ & $C$ & $n,m$ & source \\[1mm]
\hline \rule{0mm}{5mm}
111 & 0.004 & 2,$\frac{1}{2}$ & $\nu$-less $\beta\beta$ decay \\[1mm]
\hline \rule{0mm}{7mm}
$\begin{array}{c} 112 \\ 113 \end{array}$
    & 0.04    & 1,0             & CC universality \\
\hline \rule{0mm}{7mm}
$\begin{array}{c} 121 \\ 131 \end{array}$
    & 0.07   & 1,0             & atomic $P$-violation \\
\hline \rule{0mm}{8mm}
$\begin{array}{c} 122 \\ 133 \end{array}$
    & $\begin{array}{c} 0.08 \\ 0.003 \end{array}$
    & $\frac{1}{2}$,0           & $\nu_e$ mass \\
\hline \rule{0mm}{8mm}
123 & $\begin{array}{c} 0.52 \\ 0.28 \end{array}$    
    & $\begin{array}{c} 1, 0 \\ \frac{1}{2}, 0 \end{array}$   
    & $\begin{array}{c} {\rm F-B~asymmetry} \\ D-\overline{D}~{\rm
      mixing}^{\ast} \end{array}$ \\[3mm]
\hline \rule{0mm}{5mm}
132 & 0.68     & 1,0             & $R_e$ $(Z_0)$ \\[1mm]
\hline
\end{tabular}
\caption{\it Low-energy constraints on $R$-parity violating couplings
  $\lambda'_{ijk} < C \left(M_{\tilde{q}} / 200\,{\rm GeV}\right)^n
  \left(m_{\tilde{g}} / 1\,{\rm TeV}\right)^m$ relevant for $e^+p$
    scattering \protect\cite{dreiner2}. The limit on $\lambda'_{123}$
    from $D-\overline{D}$ mixing, marked by ${}^\ast$, involves quark
    mixing and is thus model-dependent.}
\label{rlimits}
\end{center} 
\end{table}
%----------------------------------

%----------------------------------
\begin{figure}[htb] 
\unitlength 1mm
\begin{picture}(160,120)
%\lattice
\put(-5,-74){
\epsfxsize=18cm
\epsfysize=23cm
\epsfbox{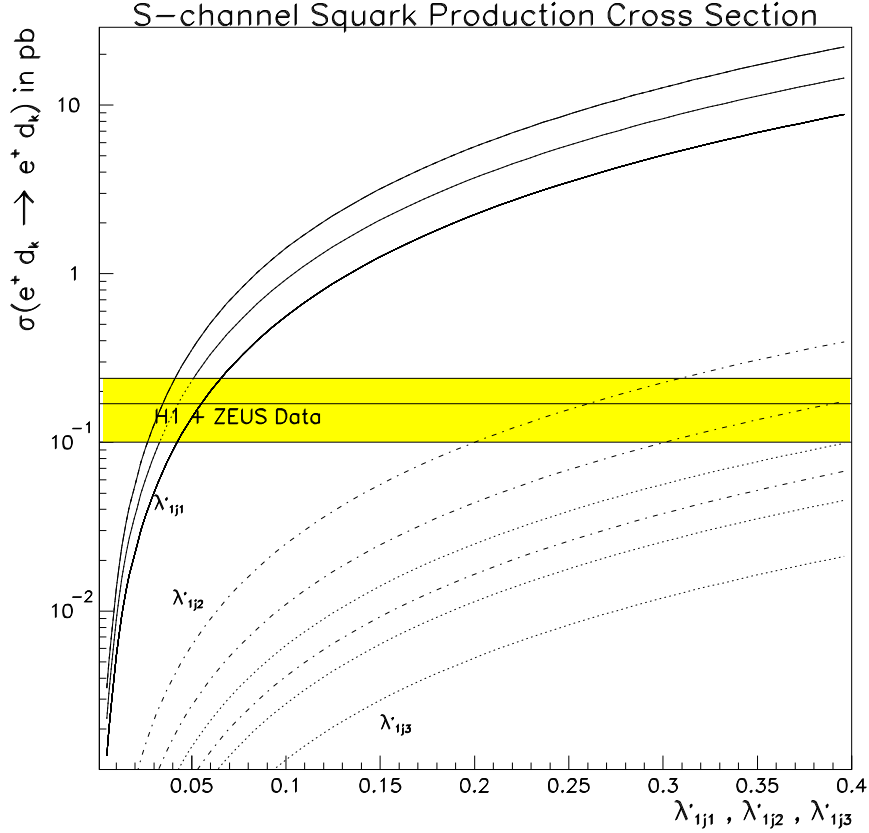}
}
\end{picture}
\caption{\it Cross section for $e^+d_k \rightarrow \tilde{u}_j
  \rightarrow e^+d_k$ as a function of the coupling $\lambda'_{1jk}$
  for $d$ valence quarks (full), $s$ quarks (dash-dotted) and $b$
  quarks (dotted) assuming $B_{eq} = 1$. The curves from top to bottom
  correspond to $M_{\tilde{u}_j} = 200$, $210$, and $220$ GeV. The
  shaded region shows the excess cross section $\sigma_{exp} =
  (0.17 \pm 0.07)$ pb from 1994 - 96 HERA data for $Q^2 > 20,000$
  GeV${}^2$ (from Ref.\ \protect\cite{dreiner}).}
\label{figdm}
\end{figure}
%----------------------------------

Expanding the superfields in (4) in terms of matter
fields, one finds interaction terms which allow for resonance
production of squarks at HERA \cite{rp}:
\begin{equation}
e^+ d_R^k \rightarrow \tilde{u}_L^j
~~~~~~ (\tilde{u}^j = \tilde{u}, \tilde{c}, \tilde{t}),
\end{equation}
\begin{equation}
e^+ \bar{u}_L^j \rightarrow \overline{\tilde{d}}_R^k
~~~~~~ (\tilde{d}^k = \tilde{d}, \tilde{s}, \tilde{b}).
\end{equation}
%----------------------------------
\begin{figure}[tbph] 
\unitlength 1mm
\begin{picture}(150,145)
%\lattice
\put(-25,-66){
\epsfxsize=19cm
\epsfysize=24cm
\epsfbox{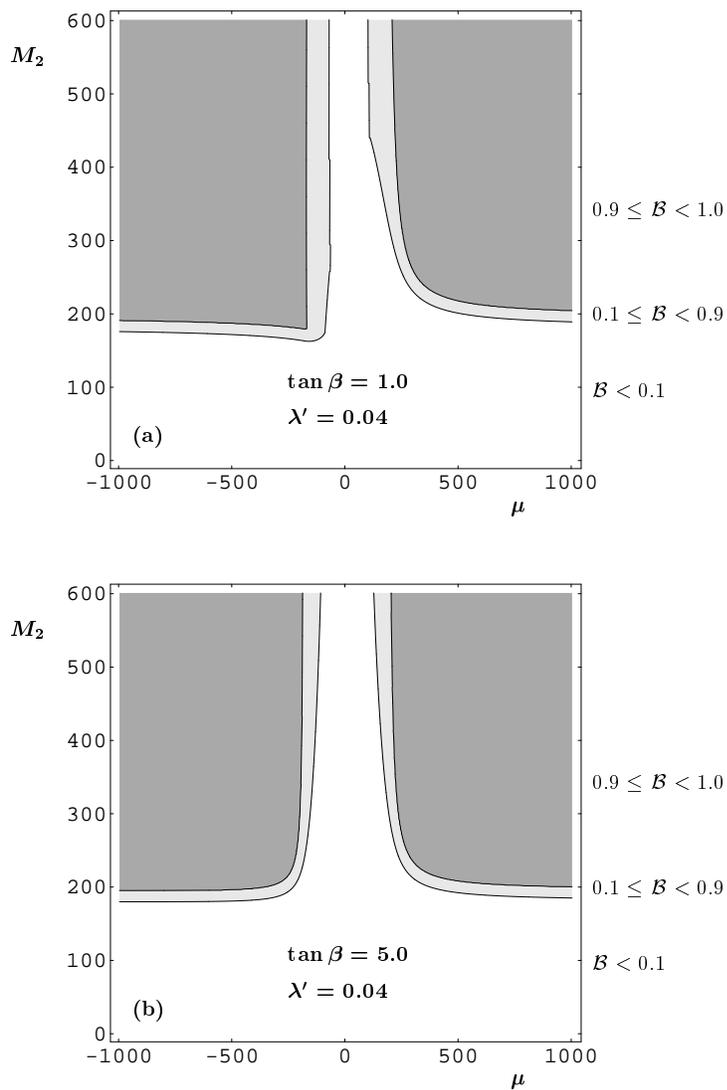}
}
\end{picture}
\caption{\it Contours of $B(\tilde{t} \rightarrow e^+d)$ in the
  $\mu-M_2$ plane assuming vanishing stop left-right mixing.  The LEP2
  bound of 85 GeV for the chargino mass is taken into account. From
  Ref.\ \protect\cite{br}).}
\label{figbr}
\end{figure}
%----------------------------------
The cross sections are determined by the coupling constants
$\lambda'_{1jk}$. Similarly as the leptoquark Yukawa couplings
$\lambda_{L,R}$ from Table \ref{tabprop}, these couplings are strongly
constrained by existing data. The relevant bounds are summarized in
Table \ref{rlimits}.  As already pointed out, since the excess of
events was observed in $e^+p$ but not in $e^-p$ scattering, the
process of class (6) involving the $\bar u$ sea is unlikely. Moreover,
the coupling strength $\lambda'_{11k} \simeq e$, required for
production off sea quarks, is incompatible with the existing bounds.
This also applies to the $e^+ \bar c$ channel with the marginal
exception of the subprocess $e^+ \bar c \to \bar{\tilde{t}}$
\cite{dreiner}. The top sea plays no role.  Turning to the processes
of class (5), one finds three possible explanations of the HERA
anomaly \cite{squarks}$^,$\footnote{For a discussion of the strange
  stop scenario see in particular Ref.\ \cite{ellis}}~~:
\begin{eqnarray}
e^+d \to \tilde{c} ~~(\lambda'_{121}),\\
e^+d \to \tilde{t} ~~(\lambda'_{131}),\\
e^+s \to \tilde{t} ~~(\lambda'_{132}).
\end{eqnarray}
The corresponding cross sections are plotted in Fig.\ \ref{figdm} for
$M_{\tilde{q}}=200$ to 220 GeV, and setting $B_{eq}=1$.  As can be
seen, within the limits on $\lambda'$ quoted in Tab.\ \ref{rlimits}
one can still afford branching ratios for $\tilde{c}, \tilde{t}
\rightarrow e^+d$ below 0.7, necessary in order to avoid the D0/CDF
mass bounds.  Studies \cite{raych,br} have shown that one can indeed
find allowed regions in the supersymmetry parameter space in which
$B_{eq} < 0.7$.  This is exemplified in Fig.\ \ref{figbr} for
$B(\tilde{t} \to e^+ d)$.  However, there is not too big a room for a
consistent squark interpretation of the HERA anomaly.

%----------------------------------
\begin{figure}[htp] 
\unitlength 1mm
\begin{picture}(155,90)
%\lattice
\put(0,0){
\epsfig{
file=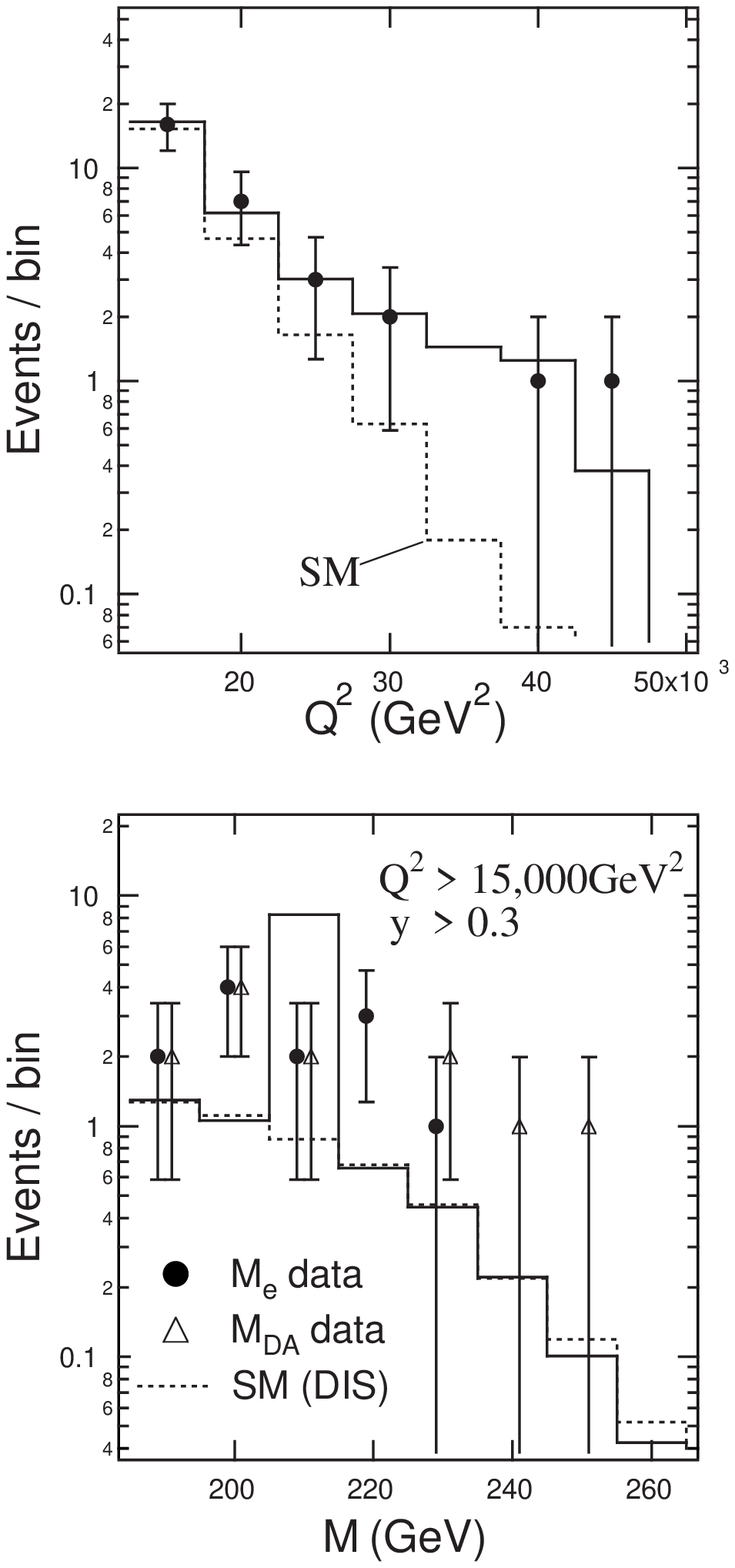,
height=8.4cm,
width=7.3cm,
bbllx=140,bblly=100,bburx=410,bbury=400,
clip
}
}
\put(76,0){
\epsfig{
file=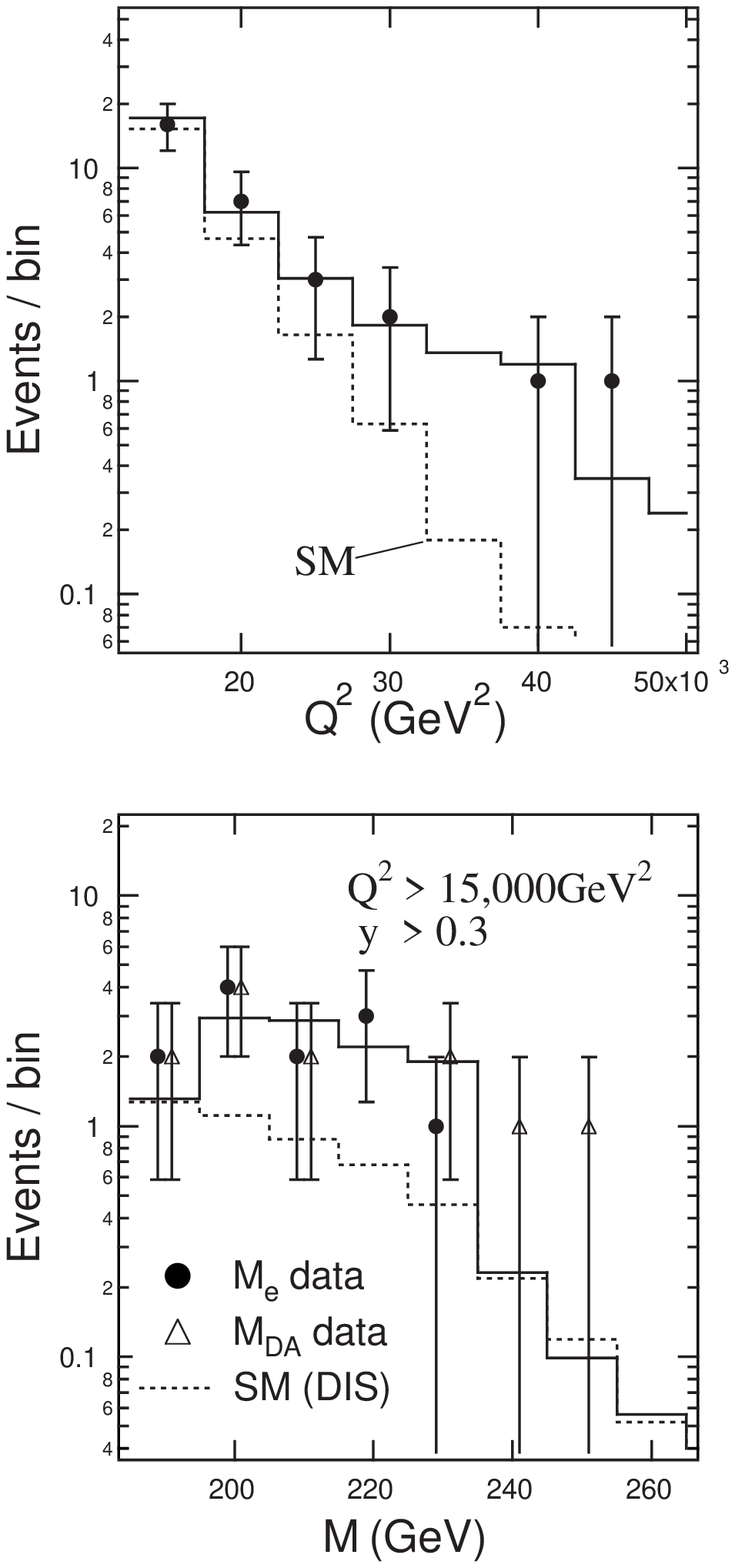,
height=8.4cm,
width=7.3cm,
bbllx=140,bblly=100,bburx=410,bbury=400,
clip
}
}
\end{picture}
\caption{\it Distributions in $M=\protect\sqrt{xs}$ of 
  the 1994 - 96 HERA data (H1 and ZEUS combined) in comparison with the
  distributions in a single stop scenario (left, $M_{\tilde{t}} = 210$
  GeV, $\lambda'_{131} = 0.04$) and in a mixed left-right stop scenario
  (right, $M_{\tilde{t}_1} = 205$ GeV, $M_{\tilde{t}_2} = 225$ GeV,
  $\theta_t = 0.95$, $\lambda'_{131} = 0.045$). From Ref.\ 
  \protect\cite{kon}.}
\label{figkon}
\end{figure}
%----------------------------------

Note that the NC events from $\tilde{t}, \tilde{c} \to e^+ d$ have the
same visible particles as the DIS-NC events. This is not expected for
the CC events originating from cascade decays of squarks on the one
hand, and DIS-CC events on the other.

Finally, the difficulty to interpret the excess of events as a
single-resonance effect may also find a reasonable solution \cite{kon}.
In the MSSM each fermion has two superpartners which mix in general. In
the case of stop this mixing may be sizeable and lead to two mass
eigenstates with a small but pronounced mass difference. Such a case is
illustrated in Fig.\ \ref{figkon}.  The resulting mass distribution can
apparently mimic a continuum effect.

%
%%%%%%%%%%%%%%%%%%%%%%%%%%%%%%%%%%%%%%%%%%%%%%%%%%%%%%%%%%%%%%%%%%%%%%
\section{Contact Interactions}

With the new 1997 HERA data it has become somewhat more likely that
the observed excess of events is due to some continuum mechanism. An
appropriate and very general description is provided by contact
interactions.  They could originate from the exchange of a new heavy
particle, or from lepton and quark substructure.  For NC lepton-quark
scattering, one may use the following effective Lagrangean:
\begin{equation}
{\cal L}_{\rm eff} = \sum_{\begin{array}{c}
                           {\tiny i,k = L,R} \\ 
                           {\tiny q = u, d, \cdots}
                           \end{array}} 
\eta^q_{ik} \frac{4\pi}{(\Lambda^q_{ik})^2}
\left(\bar{e}_i \gamma^{\mu} e_i \right)
\left(\bar{q}_k \gamma_{\mu} q_k \right)~.
\label{CI}
\end{equation}
At $Q^2 \ll \Lambda^2$, the interference of contact terms with the
standard model amplitudes leads to an enhancement or suppression of
the NC cross section depending on the sign $\eta_{ik}$.  The $x$ and
$Q^2$ dependence of the deviations is expected to be rather smooth. In
Ref.~\cite{rr} it was shown that $e^+p$ scattering is particularly
sensitive to LR and RL contact terms, while $e^-p$ scattering is
better for probing LL and RR helicity structures.  Furthermore, it was
pointed out that destructive interference leads to a rather sharp
onset of the deviations with $Q^2$.  An explanation of the HERA data
by contact terms is possible with $\Lambda$ of the order of 3 TeV
\cite{CI}.

%----------------------------------
\begin{table}[tbph]
\begin{center}
\begin{tabular}{|c|c|}\hline \rule{0mm}{5mm}
Source & Limit \\[1mm]
\hline \rule{0mm}{5mm}
Tevatron (CDF)\cite{xwu}  & $\Lambda > 2.5 \div 6.0$ TeV \\[1mm]
\hline \rule{0mm}{5mm}
HERA (H1)\cite{h1ci}      & $\Lambda > 1.0 \div 2.5$ TeV \\[1mm]
\hline \rule{0mm}{5mm}
LEP2 (OPAL)\cite{eeqq}    & $\Lambda > 1.1 \div 5.2$ TeV \\[1mm]
\hline \rule{0mm}{5mm}
atomic $P$-violation\cite{deandrea} 
                          & $\Lambda > 7.4 \div 12.3$ TeV \\[1mm]
\hline \rule{0mm}{5mm}
CCFR (for $\nu_{\mu}\nu_{\mu}qq$)\cite{gallas} 
                          & $\Lambda > 1.8 \div 7.9$ TeV \\[1mm]
\hline \rule{0mm}{5mm}
CKM unitarity\cite{agm}   & $\Lambda > 10 \div 90$ TeV \\[1mm]
\hline
\end{tabular}
\caption{\it Bounds on the scale of contact terms.}
\label{CIlimits}
\end{center} 
\end{table}
%----------------------------------

%----------------------------------
\begin{figure}[bhtp] 
\unitlength 1mm
\begin{picture}(155,113)
%\lattice
\put(0,-5){
\epsfig{
file=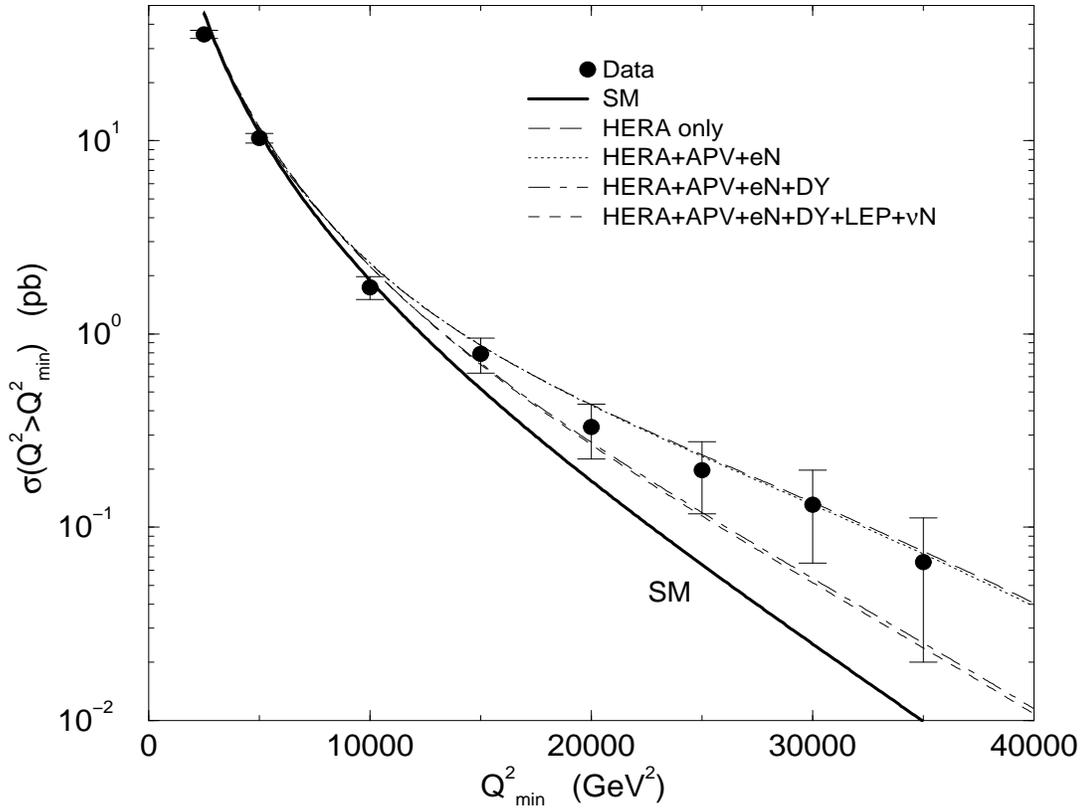,
height=12cm,
width=15cm,
bbllx=95,bblly=280,bburx=505,bbury=620,clip
}}
\end{picture}
\caption{\it Fits of contact interactions to the measured cross
  section $\sigma(Q^2 > Q^2_{min})$ for $e^+p \rightarrow e^+X$
  including bounds from low-energy and collider experiments (for
  details see Ref.\ \protect\cite{barger}).}
\label{CIfig}
\end{figure}
%----------------------------------

As is obvious from (\ref{CI}), contact interactions in $eq \rightarrow
eq$ also modify the potential in atoms and affect the crossed channels
$e^+e^- \rightarrow q\bar{q}$ and $q\bar{q} \rightarrow e^+e^-$.
Therefore, the existence of contact terms is strongly constrained by
atomic parity violation (APV) experiments, hadron production at LEP,
and Drell-Yan production at the Tevatron. Moreover, $SU(2)\times
U(1)$ symmetry implies the existence of contact interactions involving
neutrinos. Typical bounds on $\Lambda$ from the various sources are
summarized in Table \ref{CIlimits}. The existing constraints on $eeqq$
contact terms still allow a reasonable fit to the excess of NC events
as shown in Fig.\ \ref{CIfig}, provided the APV bound is avoided by
choosing an $P$-even combination of contact terms\cite{CI}. However,
an analogous explanation of the possible excess of CC events by $e\nu
q q'$ contact interactions is ruled out\cite{agm}. Finally, if this
interpretation of the signal in NC $e^+p$ scattering is correct, one
should observe a similar effect in $e^-p$ scattering.

%
%%%%%%%%%%%%%%%%%%%%%%%%%%%%%%%%%%%%%%%%%%%%%%%%%%%%%%%%%%%%%%%%%%%%%%
\section{Conclusions}

For the time being, it is an open question whether or not the excess
of high-$Q^2$ events observed at HERA is a statistical fluctuation or
a physical effect. If it is a real signal, then it very likely
originates from new physics beyond the standard model. Making this
assumption, the present data slightly favour some continuum mechanism,
but do not yet allow to rule out a resonance effect.  Both kinds of
interpretations are tightly constrained by measurements at LEP2 and
the Tevatron, as well as by low-energy data.  These bounds rule out
the simplest leptoquark scenarios and do also not leave much room for
the squark interpretation. Particularly difficult would be the
explanation of an excess of CC events.  At any rate, if the excess of
high-$Q^2$ events is confirmed by future data it is likely that
related signals will soon show up in other experiments.

%
%%%%%%%%%%%%%%%%%%%%%%%%%%%%%%%%%%%%%%%%%%%%%%%%%%%%%%%%%%%%%%%%%%%%%%
\section{Acknowledgments}
We wish to thank J.\ Kalinowski and P.\ Zerwas for the fruitful
collaboration on the subject of this talk, and T.\ Kon for providing us
with an updated version of Fig.\ \ref{figkon}. This work was supported
by the Bundesministerium f\"ur Bildung, Wissenschaft, Forschung und
Technologie, Bonn, Germany, Contracts 05 7BI92P (9) and 05 7WZ91P
(0). R.\ R.\ wants to thank B.\ Kniehl and G.\ Kramer for their kind
invitation to and hospitality at the workshop. 

%
%%%%%%%%%%%%%%%%%%%%%%%%%%%%%%%%%%%%%%%%%%%%%%%%%%%%%%%%%%%%%%%%%%%%%%

\end{document}